\documentclass[cameraready]{Interspeech}

\title{Speaker-Reasoner: Scaling Interaction Turns and Reasoning Patterns for Timestamped Speaker-Attributed ASR}

\author[affiliation={1}]{Zhennan}{Lin}
\author[affiliation={2}]{Shuai}{Wang}
\author[affiliation={1}]{Zhaokai}{Sun}
\author[affiliation={3}]{Pengyuan}{Xie}
\author[affiliation={3}]{Chuan}{Xie}
\author[affiliation={3}]{Jie}{Liu}
\author[affiliation={3}]{Qiang}{Zhang}
\author[affiliation={1}, correspondingauthor]{Lei}{Xie}

\address{
    $^1$ Audio, Speech and Language Processing Group (ASLP@NPU), School of Computer Science, Northwestern Polytechnical University \\
    $^2$ School of Intelligence Science and Technology, Nanjing University \\
    $^3$ Shanghai Lingguang Zhaxian Technology
}

\email{znlin@mail.nwpu.edu.cn, lxie@nwpu.edu.cn}


\usepackage{comment}
\usepackage{algorithm}
\usepackage{amsmath}
\usepackage{amssymb}
\usepackage{algpseudocodex}
\usepackage{multirow}
\usepackage{booktabs,colortbl}
\usepackage[table]{xcolor}

\expandafter\def\expandafter\normalsize\expandafter{%
    \normalsize
    \setlength\abovedisplayskip{3pt} 
    \setlength\belowdisplayskip{3pt} 
    \setlength\abovedisplayshortskip{0pt} 
    \setlength\belowdisplayshortskip{0pt} 
}

\begin{document}

\maketitle

\begin{abstract}

    Transcribing and understanding multi-speaker conversations requires speech recognition, speaker attribution, and timestamp localization. While speech LLMs excel at single-speaker tasks, multi-speaker scenarios remain challenging due to overlapping speech, backchannels, rapid turn-taking, and context window constraints. We propose Speaker-Reasoner, an end-to-end Speech LLM with agentic multi-turn temporal reasoning. Instead of single-pass inference, the model iteratively analyzes global audio structure, autonomously predicts temporal boundaries, and performs fine-grained segment analysis, jointly modeling speaker identity, gender, timestamps, and transcription. A speaker-aware cache further extends processing to audio exceeding the training context window. Trained with a three-stage progressive strategy, Speaker-Reasoner achieves consistent improvements over strong baselines on AliMeeting and AISHELL-4 datasets, particularly in handling overlapping speech and complex turn-taking\footnote{https://github.com/ASLP-lab/Speaker-Reasoner}.
\end{abstract}

\section{Introduction}

    \begin{figure*}[t]
      \centering
      \vspace{-10pt}  
      \includegraphics[width=.87\linewidth]{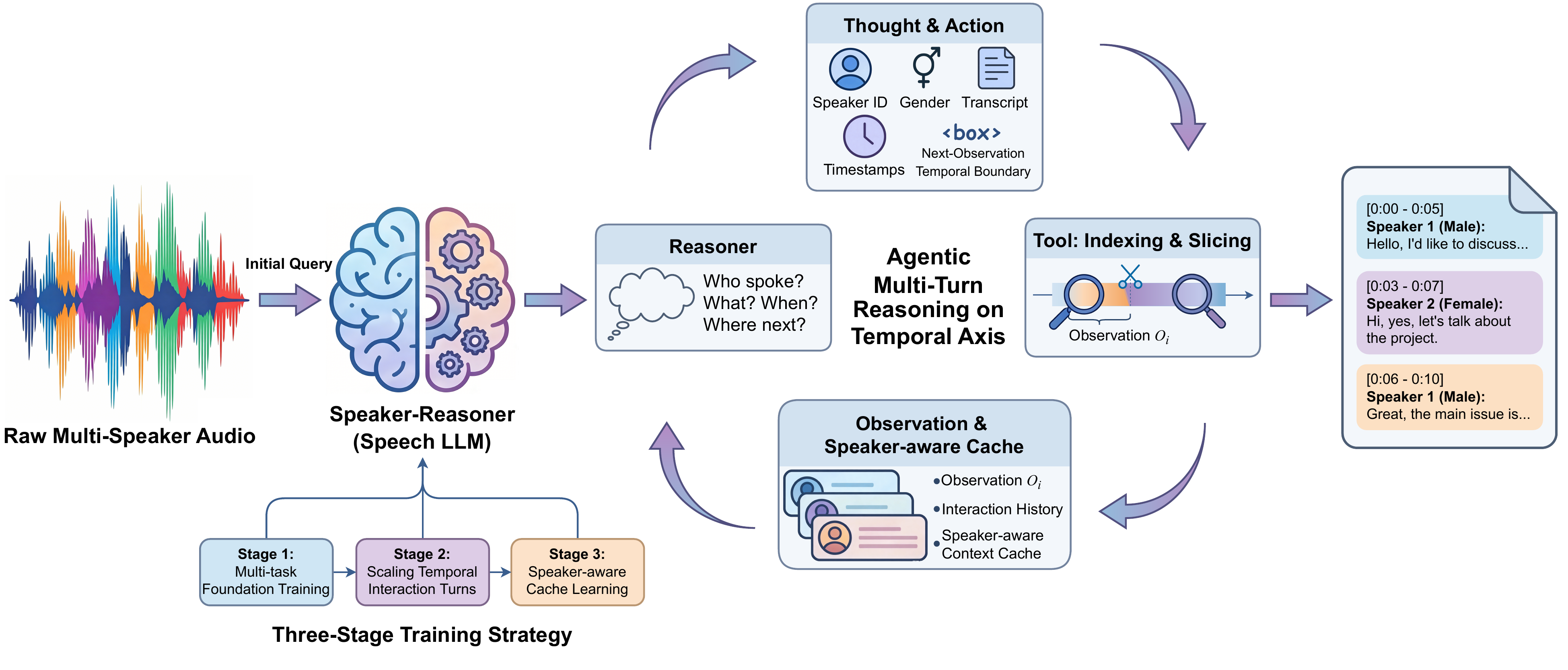}
      \setlength{\abovecaptionskip}{5pt}   
      \setlength{\belowcaptionskip}{-10pt} 
      \caption{The overview of Speaker-Reasoner. The model employs an agentic multi-turn reasoning mechanism on the temporal axis, utilizing an indexing and slicing tool and a speaker-aware context cache to iteratively generate speaker identity, gender, timestamps, and transcription from raw multi-speaker audio.}
      \label{fig:speaker_reasoner}
    \end{figure*}

   In real-world multi-speaker conversational scenarios such as meetings and phone calls, comprehensive conversation understanding requires more than speech recognition alone. It demands the joint modeling of speaker attribution, fine-grained timestamp localization, and transcription~\cite{DBLP:conf/interspeech/GaoW0000CSS25, DBLP:journals/corr/abs-2508-06372}. This task is essential for applications such as meeting transcription and intelligent assistants, yet it poses significant challenges including overlapping speech, backchannels, and rapid turn-taking, all of which degrade both speaker attribution and transcription accuracy.

    Traditional approaches adopt cascaded frameworks that pipeline independent speaker diarization (SD) and ASR modules~\cite{DBLP:conf/slt/RajDCEHH0DYLKLW21, DBLP:conf/icassp/YuZGFDZHXTWQLYM22, DBLP:conf/icassp/CornellJ0S24, DBLP:journals/taslp/BoddekerSWHR24}. While modular in design, these systems suffer from error propagation between stages, difficulty coordinating outputs in the presence of overlapping speech, and the absence of joint optimization~\cite{DBLP:journals/corr/abs-2306-13734, app15042002, DBLP:journals/corr/abs-2508-06372}. End-to-end approaches have been proposed to address these limitations, with serialized output training (SOT)~\cite{DBLP:conf/interspeech/KandaGWMY20, DBLP:conf/interspeech/LiQ0KWYQ023} emerging as the dominant paradigm. However, SOT-based models rely on single-pass sequence-to-sequence inference and lack specialized mechanisms for complex phenomena such as simultaneous speech and interruptions.

    Recent speech large language models (Speech LLMs)~\cite{DBLP:journals/corr/abs-2407-10759, DBLP:journals/corr/abs-2504-18425, DBLP:journals/corr/abs-2507-08128, DBLP:journals/corr/abs-2507-16632} such as MiMo-Audio~\cite{DBLP:journals/corr/abs-2512-23808} and Qwen3-Omni~\cite{DBLP:journals/corr/abs-2509-17765} have demonstrated strong performance across a range of speech tasks, yet these models are primarily designed for single-speaker scenarios. Recent efforts~\cite{DBLP:journals/corr/abs-2508-06372, DBLP:journals/corr/abs-2601-18184, DBLP:journals/corr/abs-2601-06896, shi2025trainshortinferlong, DBLP:conf/icassp/MengH0LWWWLM25} have begun extending Speech LLMs to multi-speaker settings, but three key limitations persist. First, task coverage remains incomplete: full timestamped speaker-attributed ASR (SA-ASR), which jointly models speaker identity, timestamps, and transcription, remains underexplored, and attributes such as speaker gender are often overlooked. Second, inference strategies remain simplistic, relying on single-pass SOT-style decoding without mechanisms tailored to overlapping speech or rapid turn-taking. Third, fixed context windows restrict processing to short segments, preventing the handling of long-form recordings common in real-world applications.

    Recent visual reasoning advances~\cite{DBLP:journals/corr/abs-2509-07969, DBLP:journals/corr/abs-2510-20579} demonstrate that complex reasoning problems can be addressed by scaling interaction turns and enriching reasoning patterns. Inspired by this, we observe that SA-ASR inherently requires progressive reasoning along the temporal axis: the model must first understand the overall audio structure as global context, then progressively localize speaker boundaries, and finally perform fine-grained segment analysis. This global-to-local paradigm naturally aligns with multi-turn interaction mechanisms. Therefore, we propose Speaker-Reasoner, an end-to-end Speech LLM for SA-ASR featuring agentic multi-turn reasoning on the temporal axis. Unlike existing methods performing single-turn sequence-to-sequence inference, Speaker-Reasoner employs iterative tool-based interactions: analyzing complete audio for global speaker distribution, autonomously predicting temporal boundaries, extracting and analyzing local slices, and reasoning about whether to continue or terminate.

    The main contributions are as follows.
(1) We propose \emph{Speaker-Reasoner}, an end-to-end Speech LLM for timestamped speaker-attributed ASR that iteratively performs global-to-local temporal inference via boundary proposal and slice-level decoding.
(2) We introduce a speaker-aware context cache to process long-form recordings beyond the model context window while preserving speaker consistency across segments.
(3) We develop a three-stage progressive training strategy to equip the model with multi-task speech understanding, temporal interaction capability, and cache usage.
(4) Experiments on AliMeeting and AISHELL-4 show consistent improvements over strong end-to-end baselines, with notable gains under overlap and rapid turn-taking.

\section{Method}
\label{sec:method}

Speaker-Reasoner addresses speaker-attributed ASR for multi-speaker long-form recordings.
The model takes raw multi-speaker audio as input and produces outputs containing speaker identity, gender, timestamps, and transcription through multi-turn interaction. 

The key challenge is that a single-pass decoder often struggles with overlapping speech and rapid turn-taking, while being constrained by fixed context windows.
We therefore factor inference into an iterative global-to-local procedure:
(1)~obtain a global speaker inventory summary,
(2)~process the audio in a sequence of temporally indexed observations, and
(3)~maintain cross-chunk speaker consistency with a speaker-aware context cache.
Figure~\ref{fig:speaker_reasoner} illustrates the overall architecture and inference flow.

\vspace{-2mm}
\subsection{Model Architecture}
\label{subsec:architecture}

Speaker-Reasoner follows an audio-to-text instruction-tuned Speech LLM paradigm,
consisting of (i)~an acoustic encoder, (ii)~a modality projector, and (iii)~an autoregressive LLM decoder.
All components are initialized from Qwen3-Omni~\cite{DBLP:journals/corr/abs-2509-17765} unless otherwise stated.

\noindent \textbf{Acoustic encoder}
Audio Transformer (AuT) encoder from Qwen3-Omni is adopted. Built on the Transformer architecture, this encoder incorporates a Conv2D module that downsamples audio features by a factor of 8, producing a compressed feature sequence at a frame rate of 12.5 Hz.

\noindent \textbf{Projector}
A lightweight two-layer MLP projector maps acoustic embeddings into the LLM embedding space, enabling alignment between acoustic tokens and text tokens.

\noindent \textbf{LLM decoder}
We use the Thinker module from Qwen3-Omni, a Mixture-of-Experts (MoE) Transformer that
activates a sparse subset of experts per token, maintaining high model capacity while ensuring computational efficiency.
The decoder receives the projected audio embeddings~$\mathbf{z}$ concatenated with text instructions
and previously generated tokens, producing responses autoregressively.

\vspace{-2mm}
\subsection{Iterative Temporal Interaction}
\label{subsec:interaction}

We perform multi-turn interaction by processing the audio incrementally along the temporal axis.
At turn~$i$, the model receives an observation~$O_i$, the interaction history~$\mathcal{H}_{i-1}$,
and the speaker-aware cache~$\mathcal{C}_{i-1}$ (Section~\ref{subsec:cache}),
and produces (i)~speaker identity, gender, timestamps, and transcription within~$O_i$ and (ii)~a boundary decision for the next observation window.

\noindent \textbf{Observation definition.}
An observation is a contiguous time slice of the original audio,
$O_i = x[t_i^{\mathrm{st}} : t_{i}^{\mathrm{ed}}]$, with $0 \le t_1^{\mathrm{st}} < t_2^{\mathrm{st}} < \cdots < t_{N+1}^{\mathrm{st}} \le T$.
The model is thus prompted with the triple $(O_i,\, \mathcal{H}_{i-1},\, \mathcal{C}_{i-1})$.

\noindent \textbf{Boundary construction and supervision.}
Training-time observation boundaries are derived from speaker-level segments to encourage the model to resolve turn-taking and partial overlaps.
Given the current observation~$O_i$, the subsequent observation~$O_{i+1}$ is defined
as the nearest speaker segment whose temporal overlap with~$O_i$ falls below a threshold.
Formally, for two segments~$A$ and~$B$ with durations~$d_A$, $d_B$ and intersection duration~$d_{AB}$, we require:
\begin{equation}
  \frac{d_{AB}}{d_A} < \tau
  \quad\text{and}\quad
  \frac{d_{AB}}{d_B} < \tau,
\end{equation}
where $\tau = 0.8$ in our experiments.
Segments exceeding this threshold are merged into the same observation for joint processing,
accommodating heavily overlapping events such as backchannels and simultaneous utterances.
At inference time, the model outputs discrete boundary tokens representing $t_{i+1}^{\mathrm{st}}$ and $t_{i+1}^{\mathrm{ed}}$, enabling fully autonomous temporal indexing.

\noindent \textbf{Interaction protocol.}
In the first turn, the model performs global analysis, outputting a summary of speaker count and gender distribution. In subsequent turns, the model processes each observation $O_i$ conditioned on the accumulated history. The interaction terminates when the model encloses the consolidated transcript within \texttt{<answer>} and \texttt{</answer>} tags, yielding a chronologically sorted speaker-attributed transcript.

\vspace{-2mm}
\subsection{Speaker-aware Context Cache}
\label{subsec:cache}

Long-form audio often exceeds the training context window, causing speaker identity drift across chunks. We introduce a speaker-aware cache that supplies speaker-specific acoustic references from previously processed content.

\noindent \textbf{Cache entry representation.}
Each cache entry is a tuple:
$c = \bigl(s, \tilde{x}[t_{\mathrm{st}}:t_{\mathrm{ed}}], \tilde{y}\bigr)$
, where $s$ is the speaker label, $\tilde{x}$ is an audio slice, and $\tilde{y}$ is the corresponding transcript.
Cache entries are prepended to the prompt as structured context.

\noindent \textbf{Cache training simulation.}
To expose the model to cache usage during training, we employ a random sampling strategy.
For each training instance, we randomly select a subset of speakers from the same session
and retrieve 1--5 segments per speaker from non-overlapping earlier audio regions,
which are prepended as cache context.
Speaker labels are reassigned by order of first appearance in the augmented input sequence,
encouraging permutation-invariant speaker assignment and ensuring label continuity
between the cached history and the current observation.

\noindent \textbf{Cache selection at inference.}
During inference, we maintain an observation buffer $B = \{O_1, \ldots, O_n\}$
of previously processed segments.
For each speaker $s \in S$, candidate segments are scored jointly by duration and recency: $\phi = d \cdot \left(1 + \alpha \cdot \frac{i}{n}\right)$,
where $d$ is the segment duration, $i$ is its position in the buffer,
and $\alpha$ controls the recency weight.
We retain the top-$k$ segments per speaker as cache entries (Algorithm~\ref{alg:cache}).
These entries provide speaker-specific acoustic references that stabilize identity assignment
for both recurring and newly emerging speakers.

\begin{algorithm}[t]
\caption{Speaker-aware Cache Selection}
\label{alg:cache}
\begin{algorithmic}[1]
\Require Observation buffer $B=\{O_1,\ldots,O_n\}$, speaker set $S$,
         recency weight $\alpha$, cache size $k$
\Ensure  Selected cache entries $C$
\State Initialize empty candidate list $\mathcal{L}_s$ for each $s \in S$
\For{$i = 1$ \textbf{to} $n$}
    \State $\rho \leftarrow i/n$ \Comment{recency ratio}
    \For{each segment $(s,\, t_{\mathrm{st}},\, t_{\mathrm{ed}},\, \text{text})$ in $O_i$}
        \If{$s \in S$}
            \State $d   \leftarrow t_{\mathrm{ed}} - t_{\mathrm{st}}$
            \State $\phi \leftarrow d \cdot (1 + \alpha \cdot \rho)$
            \State Append $(O_i,\,\phi)$ to $\mathcal{L}_s$
        \EndIf
    \EndFor
\EndFor
\For{each speaker $s \in S$}
    \State Sort $\mathcal{L}_s$ by $\phi$ in descending order
    \State $C[s] \leftarrow$ top-$k$ entries of $\mathcal{L}_s$
\EndFor
\State \Return $C$
\end{algorithmic}
\end{algorithm}

\vspace{-2mm}
\subsection{Training Objective and Curriculum}
\label{subsec:training}

We adopt a three-stage curriculum that progressively equips the model with
(i)~multi-task perception, (ii)~temporal interaction, and (iii)~long-context robustness.

\noindent \textbf{Stage~1: Multi-task foundation.}
We train the model with teacher forcing to generate the full structured output sequence
for each recording, encompassing speaker labels, gender, timestamps, and transcripts.
The training objective is a standard next-token prediction loss:
\begin{equation}
  \mathcal{L}_{\mathrm{LM}} = -\sum_{t} \log p_\theta(w_t \mid w_{<t},\,\mathbf{z}),
\end{equation}
where $w_t$ denotes the $t$-th output token.
A global summary header listing speaker count and gender distribution is prepended
to encourage global-to-local processing awareness.

\noindent \textbf{Stage~2: Temporal interaction learning.}
The model is trained to operate on observation $O_i$ conditioned on interaction history.
In addition to the language modeling loss, we supervise predicted boundary tokens:
\begin{equation}
  \mathcal{L} = \mathcal{L}_{\mathrm{LM}} + \lambda\,\mathcal{L}_{\mathrm{bd}},
\end{equation}
where $\mathcal{L}_{\mathrm{bd}}$ is a cross-entropy loss over discretized boundary bins
and $\lambda$ is a balancing weight.

\noindent \textbf{Stage~3: Speaker-aware cache learning.}
Cache-conditioned decoding is enabled by prepending sampled historical cache entries
during training, matching inference-time prompting.
This stage improves speaker consistency across long-form audio by exposing the model
to cross-chunk speaker references and enforcing label continuity.

\section{Experiments}
\subsection{Implementation Details}
We initialize Speaker-Reasoner from Qwen3-Omni, a 30B-parameter multimodal LLM with a MoE architecture that activates 3B parameters per forward pass. Training is conducted using the MS-Swift framework~\cite{DBLP:conf/aaai/ZhaoHHWMZJWAWZC25} with Megatron-LM backend on 8 NVIDIA A100 GPUs. We apply LoRA with rank 8 and scaling factor 32 to all linear layers across the audio encoder, projector, and language model. We use the AdamW~\cite{DBLP:conf/iclr/LoshchilovH19} optimizer with a learning rate of 1e-5, cosine schedule, 0.05 warmup fraction, and minimum learning rate of 1e-6. The three training stages are executed sequentially, with each stage initialized from the checkpoint of the previous stage.

\vspace{-2mm}
\subsection{Datasets}
We conduct experiments on two Mandarin multi-speaker meeting corpora: AliMeeting~\cite{DBLP:conf/asru/LiangSYLZDCXQWCLYB23} and AISHELL-4~\cite{DBLP:conf/interspeech/FuCLJKCHXWBXDC21}. AliMeeting contains approximately 120 hours of recorded meetings with 2 to 4 speakers per session, captured using an 8-channel microphone array in 15 to 30 square meter meeting rooms. AISHELL-4 comprises 211 sessions totaling around 120 hours of meeting recordings involving 4 to 8 speakers, collected with an 8-channel circular microphone array. Both datasets provide word-level timestamps, speaker identity labels, gender annotations, and manual transcriptions. They also exhibit real-world acoustic challenges including overlapping speech, reverberation, and diverse speaking styles. Following the segmentation protocol in prior work~\cite{DBLP:journals/corr/abs-2508-06372}, we use only the first microphone channel and split the original long-form recordings into segments of 40 to 50 seconds based on voice activity detection boundaries for both training and evaluation.

\vspace{-2mm}
\subsection{Evaluation Metrics}
We evaluate speaker diarization using Diarization Error Rate (DER), which measures the fraction of time incorrectly attributed to speakers, encompassing missed speech, false alarms, and speaker confusion errors. For transcription quality, we report Character Error Rate (CER). To jointly assess transcription accuracy and speaker attribution, we adopt concatenated minimum-permutation Character Error Rate (cpCER), which concatenates all utterances per speaker, finds the optimal speaker permutation, and then computes CER. We additionally report $\Delta$cp, defined as cpCER minus CER, which isolates the error contribution from speaker assignment by factoring out intrinsic transcription errors.
\definecolor{propbg}{RGB}{236, 244, 255}    
\definecolor{propbest}{RGB}{197, 218, 251}  

\begin{table*}[th]
  \caption{Comparison of speaker diarization and recognition performance on AISHELL4-Eval
           and Alimeeting-Far test sets.
           The proposed models also serve as a stage-wise ablation across training stages.
           \textbf{Bold} denotes the best result in each column.}
  \label{tab:main_results}
  \centering
  \setlength{\tabcolsep}{5pt}
  \renewcommand{\arraystretch}{1.00}
  \resizebox{\textwidth}{!}{%
  \begin{tabular}{l cccc cccc}
    \toprule
    \multirow{2}{*}{\textbf{Model}}
      & \multicolumn{4}{c}{\textbf{AISHELL4-Eval}}
      & \multicolumn{4}{c}{\textbf{Alimeeting-Far}} \\
    \cmidrule(lr){2-5}\cmidrule(lr){6-9}
      & DER$\downarrow$ & CER$\downarrow$ & cpCER$\downarrow$ & $\Delta$cp$\downarrow$
      & DER$\downarrow$ & CER$\downarrow$ & cpCER$\downarrow$ & $\Delta$cp$\downarrow$ \\
    \midrule
    %
    \multicolumn{9}{l}{\textit{Cascade Baselines}} \\[1pt]
    \hspace{1em}Pyannote3.1 + Paraformer
      & 8.10  & 19.18 & 26.24 &  7.06
      & 19.13 & 30.15 & 45.39 & 15.24 \\\midrule
    %
    \addlinespace[4pt]
    \multicolumn{9}{l}{\textit{End-to-End Baselines}} \\[1pt]
    \hspace{1em}Gemini-2.5-Pro$^{\dagger}$
      & 36.07 & 19.81 & 25.11 &  5.30
      & 56.39 & 30.16 & 39.29 &  9.13 \\
    \hspace{1em}Qwen3-Omni-30B-A3B-Instruct
      & 32.42 & 14.46 & 22.22 &  7.76
      & 37.15 & 25.40 & 36.28 & 10.88 \\
    \hspace{1em}Qwen2.5-Omni-7B
      & 85.68 & 33.37 & 60.45 & 27.08
      & 91.77 & 38.13 & 73.38 & 35.25 \\
    \hspace{1em}SpeakerLM (212.25h)
      & {--}  & 17.75 & 26.14 &  8.39
      & {--}  & 18.63 & 32.22 & 13.59 \\
    \hspace{1em}SpeakerLM (7638.95h)
      & {--}  & 17.17 & 18.37 &  1.20
      & {--}  & \textbf{13.97} & \textbf{16.05} & 2.08 \\
    \hspace{1em}VibeVoice-ASR
      & 10.88 & 22.30 & 26.30 &  4.00
      & 20.70 & 34.67 & 40.54 &  5.87 \\
    \hspace{1em}TagSpeech-Alimeeting
      & 37.51 & 35.70 & 53.44 & 17.74
      & 52.46 & 47.11 & 68.74 & 21.63 \\ \midrule
    %
    \addlinespace[4pt]
    \multicolumn{9}{l}{\textit{Ours (Stage-wise Ablation)}} \\[1pt]
    \rowcolor{propbg}
    \hspace{1em}Qwen3-Omni + SOT sft
      & {--}  & 17.65 & 19.59 & 1.94
      & {--}  & 24.24 & 26.03 & 1.79 \\
    \rowcolor{propbg}
    \hspace{1em}Speaker-Reasoner Base (Stage~1)
      & 6.24  & 14.04 & 16.54 & 2.50
      & 8.96  & 21.16 & 22.64 & 1.48 \\
    \rowcolor{propbg}
    \hspace{1em}Speaker-Reasoner Multi-turn (Stage~2)
      & \textbf{5.19} & \textbf{13.83} & 14.93 & 1.10
      & 7.47 & 20.34 & 20.29 & -0.05 \\
    \rowcolor{propbest}
    \hspace{1em}\textbf{Speaker-Reasoner Multi-turn w/ SAC (Stage~3)}
      & 5.26 & \textbf{13.83} & \textbf{14.73} & \textbf{0.90}
      & \textbf{7.34} & 20.57 & 20.43 & -0.14 \\
    \arrayrulecolor{gray!100}\hline\arrayrulecolor{black}
    \rowcolor{propbg}
    \hspace{1em}{\small Speaker-Reasoner Base 7B}
      & {\small 12.00} & {\small 15.65} & {\small 25.60} & {\small 9.95}
      & {\small 18.43} & {\small 24.97} & {\small 38.12} & {\small 13.15} \\
    \rowcolor{propbg}
    \hspace{1em}{\small Speaker-Reasoner Multi-turn 7B}
      & {\small 9.38} & {\small 15.31} & {\small 22.91} & {\small 7.60}
      & {\small 15.56} & {\small 24.33} & {\small 34.81} & {\small 10.48} \\
    \bottomrule
    \multicolumn{9}{l}{%
      \footnotesize $^{\dagger}$~Closed-source model.\quad
      DER unavailable for SpeakerLM and SOT-based models due to incompatible output formats.} \\
  \end{tabular}%
  }
\end{table*}

\vspace{-2mm}
\subsection{Performance Comparison with Baselines}

    Table~\ref{tab:main_results} compares Speaker-Reasoner against competitive baselines on AISHELL4-Eval and Alimeeting-Far test sets. Our models achieve state-of-the-art performance across all metrics. Specifically, Speaker-Reasoner Multi-turn w/ SAC achieves a DER of 5.26\% and cpCER of 14.73\% on AISHELL4, significantly outperforming the closed-source Gemini-2.5-Pro. On the more challenging Alimeeting-Far, our model maintains superiority with a DER of 7.34\% and cpCER of 20.43\%, effectively handling complex speaker overlaps.

    A notable observation is the $\Delta$cp of -0.14\% achieved by our best model on Alimeeting-Far. While $\Delta$cp is typically positive due to speaker attribution errors, a negative value indicates that cpCER is slightly lower than standard CER. This occurs because the concatenated minimum-permutation alignment in cpCER calculation can compensate for minor transcription inaccuracies when the model achieves high speaker tracking precision.

    We further validate the architecture by evaluating 7B-parameter variants. Speaker-Reasoner Multi-turn 7B outperforms specialized baselines such as VibeVoice-ASR and SpeakerLM (7638.95h) that utilize significantly more data. Our 7B model achieves cpCER of 22.91\% and 34.81\% on the two sets respectively, substantially improving over zero-shot Qwen2.5-Omni-7B. These results demonstrate that our three-stage training and multi-turn reasoning enable superior timestamped speaker-attributed ASR performance even under constrained data and parameter scales.

\vspace{-2mm}
\subsection{Ablation on Training Stages}

    \begin{table}[t]
        \setlength{\abovecaptionskip}{0pt}   
        \setlength{\belowcaptionskip}{-10pt}  
        \caption{Performance on long-form AISHELL4-Eval without segmentation.}
        \label{tab:long_audio}
        \centering
        \renewcommand{\arraystretch}{0.9}
        \begin{tabular}{l cc}
        \toprule
        \multirow{2}{*}{\textbf{Model}} & \multicolumn{2}{c}{\textbf{AISHELL4-Eval}} \\
        \cmidrule(lr){2-3}
        & DER & cpCER \\
        \midrule
        Gemini-2.5-Pro & \textbf{15.32} & \textbf{31.59} \\
        Speaker-Reasoner Multi-turn w/ SAC & 21.60 & 36.20 \\
        \bottomrule
        \end{tabular}
        \vspace{-15pt}  
    \end{table}

    Table~\ref{tab:main_results} illustrates performance evolution across the three training stages. The transition from base to multi-turn interaction yields substantial improvements, with reductions in DER, cpCER, and $\Delta$cp across both evaluation sets. This confirms that agentic reasoning effectively captures complex temporal dependencies and speaker transitions. Furthermore, the stability of DER and cpCER upon integrating SAC demonstrates that the cache extends model capabilities without compromising diarization or recognition accuracy.

\vspace{-2mm}
\subsection{Performance on Long-form Audio}
    Table~\ref{tab:long_audio} evaluates Speaker-Reasoner on long-form AISHELL4-Eval recordings without manual segmentation. Speaker-Reasoner Multi-turn w/ SAC achieves DER of 21.60\% and cpCER of 36.20\%. Although Gemini-2.5-Pro maintains a lead in these metrics, the results indicate that the speaker-aware cache effectively enables the model to maintain context across extended durations.

\vspace{-2mm}
\subsection{Evaluation of Speaker Attributes}

    \begin{table}[t]
        \setlength{\abovecaptionskip}{0pt}   
        \setlength{\belowcaptionskip}{-10pt}  
        \caption{Gender prediction accuracy (ACC) and speaker count accuracy (SCA) on AISHELL4-Eval. SCA measures whether the predicted number of speakers matches the ground truth.}
        \label{tab:acc_sca}
        \centering
        \renewcommand{\arraystretch}{0.9}
        \begin{tabular}{l cc}
        \toprule
        \textbf{Model} & \textbf{ACC} & \textbf{SCA} \\
        \midrule
        Gemini-2.5-Pro & 94.80 & 67.03 \\
        Qwen3-Omni-30B-A3B-Instruct & 97.12 & 60.49 \\
        Speaker-Reasoner Multi-turn & \textbf{96.80} & \textbf{69.03} \\
        \bottomrule
        \end{tabular}
        \vspace{-15pt}  
    \end{table}

    Table~\ref{tab:acc_sca} evaluates gender prediction accuracy (ACC) and speaker count accuracy (SCA) on AISHELL4-Eval. Speaker-Reasoner Multi-turn achieves gender accuracy of 96.80\% and SCA of 69.03\%, outperforming Gemini-2.5-Pro and zero-shot Qwen3-Omni. These results indicate that iterative global-to-local reasoning enables more reliable speaker distinction in complex multi-speaker audio, thereby improving global speaker composition estimation.

\section{Conclusion}
    In this work, we present Speaker-Reasoner, an end-to-end Speech LLM for timestamped speaker-attributed ASR. We introduce an agentic multi-turn reasoning mechanism that shifts inference from single-pass decoding to iterative global-to-local reasoning. This enables the model to autonomously resolve complex multi-speaker scenarios, while a speaker-aware cache ensures speaker consistency across long-form recordings. 
    Experiments on AliMeeting and AISHELL-4 demonstrate superior performance, suggesting the potential of scaling interaction turns and reasoning patterns for conversation understanding in speech LLMs.

\bibliographystyle{IEEEtran}
\bibliography{mybib}

\end{document}